# A STATISTICAL FRAMEWORK FOR THE ANALYSIS OF MICROARRAY PROBE-LEVEL DATA

By Zhijin Wu[1] and Rafael A. Irizarry[2]

*Brown University and Johns Hopkins University*

In microarray technology, a number of critical steps are required to convert the raw measurements into the data relied upon by biologists and clinicians. These data manipulations, referred to as *preprocessing*, influence the quality of the ultimate measurements and studies that rely upon them. Standard operating procedure for microarray researchers is to use preprocessed data as the starting point for the statistical analyses that produce reported results. This has prevented many researchers from carefully considering their choice of preprocessing methodology. Furthermore, the fact that the preprocessing step affects the stochastic properties of the final statistical summaries is often ignored. In this paper we propose a statistical framework that permits the integration of preprocessing into the standard statistical analysis flow of microarray data. This general framework is relevant in many microarray platforms and motivates targeted analysis methods for specific applications. We demonstrate its usefulness by applying the idea in three different applications of the technology.

**1. Introduction.** Microarray technology measures the quantity of nucleic acid molecules present in a biological sample referred to as the *target*. To do this, we take advantage of hybridization properties of nucleic acid and use complementary molecules immobilized on a solid surface, referred to as *probes*. The nucleic acid molecules in the target sample are labeled with fluorescent dyes and hybridized to the probes. A specialized scanner is then used to measure the amount of fluorescence at each probe which is reported as an intensity. A defining characteristic of microarray technology is that it includes thousands of probes on a relatively small surface such as a glass slide. Various manufacturers provide a large assortment of different platforms.

---

Received December 2006; revised April 2007.

[1]Supported in part by P30-AI42853.

[2]Supported in part by RO1-HG02432.

In this paper we refer to the probe hybridizing to its target phenomenon as *specific binding*. We define this term because, in practice, probes hybridize to the noncomplementary molecules as well. We refer to this phenomenon as *nonspecific binding*. Observed intensities are a combination of optical noise, a nonspecific binding component and a specific-binding component. Most manufacturers try to measure the optical and nonspecific components directly by including additional features or obtaining readings from areas of the chip with no probe. We refer to the intensities read for each of these features as *probe-level* data. In practice, various sources of variation need to be accounted for and these data are heavily manipulated before one obtains the genomic-level measurements that most biologists and clinicians use in their research. This procedure is commonly referred to as *preprocessing*. Most preprocessing procedures only provide point estimates for each genomic unit per sample. Although some methods provide measures of uncertainty for these estimates, they are often discarded in higher level analysis [Rattray et al. (2006)]. In this paper we introduce a statistical framework that permits the integration of preprocessing into the standard statistical analysis of microarray data. This general framework can be adapted to a variety of biological applications of microarray technology. We present the framework in Section 3 and demonstrate its use in Section 4 with three examples.

For the statistical framework presented in this paper it is convenient to divide the many different available platforms into two main classes. These are differentiated by the type of data they produce. We refer to the platforms that produce one set of probe-level data per microarray with some probes designed to measure specific binding and others to measure nonspecific binding as *high density oligonucleotide* platforms. Affymetrix™ GeneChip® arrays are by far the most popular manufacturer of this technology. The *two-color* platforms produce two sets of probe-level data per microarray (the red and green channels) and local optical noise levels are measured from parts of the glass slide containing no probes. No single company or academic lab dominates this market. Notice that there are a handful of platforms that do not fall into either of these categories. However, the vast majority of data produced in the past 5 years do.

The most popular microarray application of both platforms is measuring genome-wide transcription activity. In this application each gene is represented by one or more probes that will hybridize with the RNA transcribed from that gene. In practice, researchers using microarrays for this purpose start out with the probe-level data. However, most microarray products come equipped with softwares that preprocess these data into higher level measurements where each gene gets assigned one value on each array. This value is presented as the starting point for analyses that eventually lead to the results published in the scientific literature. Examples of these higher level analyses are identifying differentially expressed genes, class discovery and class prediction. In some cases, the data manipulations performed in the preprocessing step turn out to be rather complicated. Three steps typically carried out in preprocessing are:

1. Adjusting probe intensities for optical noise and/or nonspecific binding. This task is referred to as *background correction*.
2. Adjusting probe intensities to remove systematic biases due to technical variations such as different labeling efficiency, scanner setting or physical problems with the arrays. This task is referred to as *normalization*. Note that this does not necessarily mean that we transform the data to have a normal distribution, as the term traditionally implies in statistics.
3. When multiple probes represent a gene, summarizing the observed intensities to attain one number for each gene. We will refer to this step as *summarization*.

We will refer to these as the *three main preprocessing tasks*. For both platform classes, many different approaches have been proposed for each of these three steps, resulting in competing preprocessing algorithms. Most of these preprocessing algorithms do not try to estimate the measures of uncertainty that accompany the resulting gene-level expression estimates. For example, nonlinear normalization routines, such as quantile normalization [Amaratunga and Cabrera (2001)] and "variance stabilizing normalization" [Huber et al. (2002)], can artificially reduce variation of the gene-level measurements. This fact is rarely taken into account in the higher level analyses. Notice that, for researchers with the luxury of numerous array replicates, this is not necessarily a problem because measures of uncertainty can be estimated from the gene-level data. However, this situation is not common in academia and governmental institutions. Thus, for most microarray experiments, it becomes important to obtain as much information as possible about the stochastic properties of the final summary statistics from the probe-level data. By posing models for these data, any manipulation could be described statistically and bottom line results can be better understood.

Microarrays are now being used to measure genomic endpoints other than gene expression, including yeast mutant representations, the presence of Single Nucleotide Polymorphisms (SNPs), presence of deletions/insertions, and protein binding sites by chromatin immunoprecipitation (known as ChIP-chip). In each case, the units of measurement continue to be the probes. Without appropriate understanding of the bias and variance of these measurements, biological inferences based upon probe analysis will be compromised. In Section 3 we present a general statistical framework which consists of a stochastic model for probe-level data, useful for any microarray application, and procedures for quantifying answers of scientific interest that permit measuring the statistical properties introduced by the three main preprocessing tasks. In Section 4 we give examples of the usefulness of our proposal in three specific applications of microarray technology: detecting expressed genes, estimation of differential expression, and identification of synthetic lethality and fitness defects in yeast mutants. Data used in the first two examples are from a high-density oligonucleotide platform and data used in the third example are from a two-color platform.

**2. Previous work.** Various research groups have demonstrated that statistical methodology can provide great improvements over the ad-hoc preprocessing procedures offered as defaults by the companies producing the arrays. The implementation of these methods have resulted in useful preprocessing algorithms which have already provided better scientific results for users of gene expression arrays. Most of these procedures perform all three main preprocessing tasks. However, some approaches follow a step-by-step/modular approach, and others follow a global/unified approach.

For a detailed description of the three major preprocessing tasks and a review of some of the most popular preprocessing methodologies, we refer the readers to our working paper [Wu and Irizarry (2005)]. Here we describe the additive-background-multiplicative-error (addimult) model that has been implicitly or explicitly assumed to motivate most of the widely used preprocessing procedures.

2.1. *The addimult model.* After target RNA samples are prepared, labeled and hybridized with arrays, these are scanned and images are produced and processed to obtain an intensity value for each probe. These intensities represent the amount of hybridization for each probe. However, part of the hybridization is nonspecific and the intensities are affected by optical noise. Therefore, the observed intensities need to be adjusted to give accurate measurements of specific hybridization. In this paper we refer to the part of the observed intensity due to optical noise and nonspecific binding as *background noise.* Wu et al. (2004) describe experiments useful for understanding background noise behavior that empirically confirm that its effect is additive and its distribution has nonzero mean.

The component of the observed intensities related to specific binding is also affected by probe properties as well as measurement error. By using the log-scale transformation before analyzing microarray data, many investigators have implicitly assumed a multiplicative measurement error model [Dudoit et al. (2002), Kerr, Martin and Churchill (2000), Newton et al. (2001), Wolfinger et al. (2001)]. Furthermore, various groups, for example, Li and Wong (2001), have demonstrated the existence of strong multiplicative probe effects on the ability to measure specific signals.

Most ad-hoc preprocessing algorithms subtract background and then take the log which arguably implies an addimult model. However, Cui et al. (2003), Durbin et al. (2002), Huber et al. (2002) and Irizarry et al. (2003a) have explicitly proposed addimult models and motivated algorithms based on these. A general form of this model is simply

$$Y = B + S,$$
(1)

with $Y$ the observed intensity, $B$ the background noise component and $S$ the specific binding component which includes multiplicative effects.

2.2. *Modular versus unified approaches.* The three main preprocessing tasks can be performed sequentially and produce gene level measures for each gene on each array. Further higher level analyses use the summaries from preprocessing as input data. We refer to this type of data flow as the modular approach. One disadvantage of the modular approach is that often the effect of the preprocessing steps on the stochastic properties of the final statistical summaries is ignored. For example, the gene expression measures produced by preprocessing procedures often have different uncertainties. Not all preprocessing methods provide a measure of this uncertainty. Even when an uncertainty measure is provided, most high level analysis methods do not make use of it. There are some exceptions, for example, Liu et al. (2006) and Rattray (2006) show that propagating the uncertainty in gene expression measures can improve accuracy in detecting differential gene expression and principle component analysis. They still use a modular approach, but include the variance obtained from preprocessing as part of variance of the log scale expression level in high level analysis.

Another disadvantage of the stepwise modular approach is that each step is independently optimized without considering the effect of previous or subsequent steps. This could lead to sub-optimal bottom-line results. Various investigators have used the addimult model to combine the background adjustment and normalization step into a unified estimation procedure. For example, Durbin et al. (2002), Huber et al. (2002), Geller et al. (2003) and Cui et al. (2003) use addimult models to motivate a transformation of the data that removes the dependence of the variance on the mean intensity levels. However, these procedures do not define and estimate parameters that represent quantities related to a scientific question as we wish to accomplish with our general framework.

Some methods have been proposed to estimate, or test for, differential expression as part of a more general estimation procedure that performs some of the main preprocessing tasks. For example, Kerr et al. (2000) propose the use of ANOVA models to test for differential expression across different populations in two-color arrays. Their models include parameters to account for the need for normalization. However, the background adjustment step is performed separately. Wolfinger et al. (2001) propose a similar model that permits some of the effects to be random. This group developed the equivalent approach for high-density oligonucleotide arrays [Chu, Weir and Wolfinger (2002)]. In both approaches no background adjustment is performed. Hein et al. (2005) propose a Bayesian model for high-density oligonucleotide arrays that combines background adjustment and summarization, and permits the possibility of estimating more meaningful parameters along with credibility intervals. However, the normalization task is not addressed and probe effects are not considered in the summarization.

In the next section we propose a statistical framework that will permit us to estimate parameters of interest and perform all three main preprocessing tasks in one estimation procedure. The measures of uncertainty will therefore account for the preprocessing.

**3. A general statistical framework.** The first step in our proposed framework is the definition of target DNA/RNA molecule of interest. For example, in expression arrays, we are interested in RNA transcripts. Then, for each target molecule, a set of probes, that will provide specific binding measurements for this target, are identified. Probes that provide information about nonspecific binding are also identified. Finally, answers to scientific questions related to these target molecules can be quantified as summaries of the parameters in the following statistical model:

$$(2) \qquad Y_{gij}^h = O_{gij}^h + N_{gij}^h + S_{gij}^h,$$

with $g = 1, \ldots, G, i = 1, \ldots, I, j = 1, \ldots, J_g$ and $h = 1, \ldots, H$.

Here $Y_{gij}^h$ is the probe intensity read from a probe of type $h$, for target molecule $g$, in array $i$, and probe $j$. For example, in GeneChip arrays $h = 1, 2$ will correspond to PM or MM and in two-color arrays to $Red$ or $Green$. The target molecules of interest, such as mRNA transcripts or SNP sites, are indexed with $g$. The different probes used to represent a target are denoted with $j$. In many cases, for example, most two-color platforms, only one probe is used and $j$ can be omitted.

The probe intensity contains three major components: optical noise $O$, intensities due to nonspecific and specific binding, $N$ and $S$. In our model $O$ can depend on the various indexes, but in all our examples we consider it to be a constant for each array. The $N$ and $S$ components can be further decomposed into

$$(3) \qquad \begin{aligned} N_{gij}^h &= \exp(\mu_{gij}^h + \xi_{gij}^h) \quad \text{and} \\ S_{gij}^h &= \exp(\nu_i^h + \theta_{gi}^h + \phi_{gij}^h + \varepsilon_{gij}^h), \qquad \text{if } S_{gij}^h > 0. \end{aligned}$$

The mean level of nonspecific intensity for the $j$th probe of type $h$ related to target molecule $g$ is represented by $\mu_{gj}^h$, and random effects that explain differences from array to array are denoted with $\xi_{gij}^h$. The fact that the $N$ is strictly positive explains the need for background adjustment. If target molecule $g$ is present, then the specific binding component $S_g$ is formed by an array specific constant $\nu$ that explains the need for normalization, a log-scale probe effect $\phi$, measurement error $\varepsilon$ and a quantity proportional to the amount of transcript $\exp\{\theta\}$. For example, in two-color arrays, $\theta^{\text{red}}$ and $\theta^{\text{green}}$ represent the specific binding in the two channels. In GeneChip arrays, $\theta^{PM}$ represent the specific binding on the PM probe, and $\theta^{MM}$ represent the intensity due to binding of the PM target on the MM probe. It has been observed that, at least for some probes, $\theta^{MM} > 0$ [Wu and Irizarry (2004)].

The distribution of stochastic components in (3) will depend on the platform and application. However, we model $\xi$ with a normal distribution in the examples presented in this paper. Using an experiment designed specifically to motivate a stochastic model for background noise, Wu et al. (2004) demonstrate this is a reasonable assumption for GeneChip arrays. Below we present evidence that the log-normal assumption applies to two-color platforms as well. If we remove outliers, the normal assumption appears to be useful to model $\varepsilon$ as well.

Notice that some of the models motivating the unified preprocessing algorithms described in Section 2 are special cases of model (3). An example is the model proposed by Durbin et al. (2002) for two-color platforms. To obtain their model from ours, we need to assume $N$ is 0 and that $O$ is normally distributed. Instead of estimating $\theta$, Durbin et al. (2002) derive a transformation $t$ for which the variance of $\Delta = t(Y^R) - t(Y^G)$ does not depend on the expectation of $S^R$ and $S^G$. The difference $\Delta$ is used as a measure of relative expression on the two samples. Huber et al. (2002) follow a similar approach. Unlike Durbin et al. (2002), they explicitly include $\phi$ and $\nu$ in their model. As in Durbin et al. (2002), they consider $N + O$ to be normally distributed. Because their procedure was originally developed for two-color arrays, $\phi$ is absorbed into $\theta$. Using an ad-hoc robust version of maximum likelihood estimation, the parameters are estimated to derive a transformation similar to the one proposed by Durbin et al. (2002). The model described by Kerr et al. (2000) is also a special case of ours. They assume $Y$ has been background adjusted and, therefore, that $O$ and $N$ are 0. They incorporate the estimation of differential expression with the normalization step by permitting $\theta^h_{gi}$ to be constant for measurements from the same population. Hein et al. (2005) use our model as well, but impose further assumptions on the distribution of the parameters. The $\nu$ and $\phi$ parameters are not accounted for though.

The approaches described by Durbin et al. (2002) and Huber et al. (2002) assume $O + N$ to be IID normal for each hybridization. As mentioned, empirical evidence suggests that this assumption is incorrect and that the distribution of the background component is heavily right skewed. Therefore, a log-normality assumption is more appropriate. This incorrect assumption of normality has a relatively large impact on the accuracy of the expression level estimate. Figure 1 compares the resulting expression estimates ob-tained from using *VSN* procedures proposed by Huber et al. (2002) to the *GCRMA* procedure which uses a log-normal assumption [Wu et al. (2004)], using data from an assessment experiment (described in more detail in Section 4.1.3). The result from the generalized-log (*glog*) proposed by Durbin et al. (2002) is almost indistinguishable from VSN and is omitted. We also compare these to a procedure that is just like GCRMA, but does no background correction at all. The figure shows averaged log (base 2) expression estimates plotted against known log (base 2) concentration levels for data from an assessment experiment (described in more detail in Section 4.1.3). Appropriate background adjustment will yield a straight line and, according to our model, no background adjustment will yield flat local slopes for low concentrations. Notice that the procedures using the normality assumption are almost equivalent to not correcting for background.

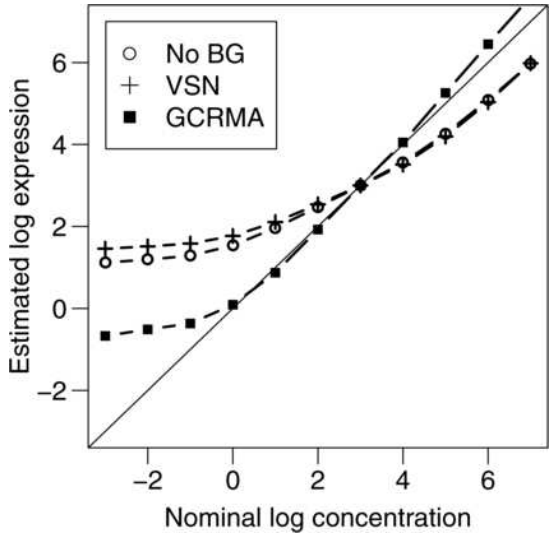

Fig. 1. *Log (base 2) expression estimates plotted against nominal log (base 2) concentration in picoMolar, computed with background adjustment described in the text. To make the curves comparable, the lines are shifted so that they have the same expression at log concentration 8 picoMolar (3 in log base 2).*

Although the proposal of using a log-normal distribution for the background noise provides great practical improvements, the major advantage of our statistical framework is that it will permit us to describe final results of scientific interest with rigorous statistical statements. We will be able to quantify scientific questions as an exercise of optimizing estimation of a set of model parameters. With the proper model in place, fitting the model will produce direct estimates of the parameters of interest along with uncertainty measures that take into account the effects of the three main preprocessing tasks.

Fitting model (3) in practice will sometimes be challenging. Many parameters in model (3) are not identifiable without certain constraints. However, the platform designs usually impose constraints that allow the parameters to be identified. For example, in GeneChip arrays we will assume that the probe-effects $\phi_{gij}^h$ does not depend on array $i$ and that $\nu_i^h$ does not depend on the probe-type $h$. In two-color platforms we will assume that $\phi_{gij}^h$ does not depend on probe-type $h$. In platforms that use isothermal design [Wang et al. (2006b)] to minimize the range of optimal hybridization temperature of all probes, it is possible to omit $\phi_{gij}^h$. Other application specific assumptions that make the model more parsimonious will be demonstrated by examples in Section 4.

The choice of which components in the model are random and which are fixed will also vary from application to application. In some applications we may model $\phi_{gi}$ with a normal distribution that does not depend on $i$ or $g$ as done by Wolfinger et al. (2001). In cases where we assume the variance of $\varepsilon$ depends on $g$, then assuming this variance follows, for example, a gamma distribution across $g$ will add power to the analysis. For gene-level data, these types of hierarchical models have greatly improved results in practice, such as the ability of finding differential expressed genes. For example, see Lonnstedt and Speed (2002), Smyth (2004), Gottardo et al. (2003), Pan et al. (2003) and Kendziorski et al. (2003). However, in spite of the improvements and the easy interpretation of the results (e.g., posterior probability of differential expression), we often find Bayesian inference computationally expensive, as observed by other researchers [Liu et al. (2006)]. In the cases when the cost of computing becomes impractical, ad-hoc versions are possible. In these cases we can still use the model assumptions to describe the statistical characteristics of the resulting data summaries.

**4. Applications.** In this section we describe how our framework can be adapted to give solutions to three important practical problems: detecting expressed genes, estimation of differential expression, and identification of synthetic lethality and fitness defects in yeast mutants. In each section we briefly describe the scientific problem, the way our framework will be implemented, a dataset used to assess the performance of our approach, and results comparing our approach to standard ones.

### 4.1. *Detecting expressed genes.*

4.1.1. *Scientific problem.* For any given target sample, it is not likely that transcripts from all genes are present. Determining which transcripts are present is sometimes of scientific interest. The Affymetrix default software (MAS 5.0) includes an algorithm for the detection of expressed genes using GeneChip arrays. The results are summarized as *detection calls* that can take the values absent (A), marginal (M) and present (P). Using our framework, one can construct an algorithm by viewing the problem as testing the hypothesis

$$\mathrm{E}[S_{gij}^{PM}] = 0 \qquad \text{for all } j = 1, \ldots, J_g,$$

for each gene $g$ on each array $i$, with the alternative hypothesis $\mathrm{E}[S_{gij}^{PM}] > 0$ (remember $S_{gij}^{PM} \geq 0$ by definition). For this application we also assume that $O$ is constant within array. Because the variability of the $\mu_{gij}^{PM}$ across $(g, j)$ has been demonstrated to be very large [Wu et al. (2004)], this problem is not trivial. The solution offered by Affymetrix can be derived by assuming $\mu_{gij}^{PM} = \mu_{gij}^{MM}$ and $S_{gij}^{MM} = 0$ for all probes [Hubbell, Liu and Mei (2002) and Liu et al. (2002)]. Under these assumptions, $E[Y_{gij}^{PM} - Y_{gij}^{MM}] = 0$ under the null hypothesis. A Wilcoxon test on $R_g = (Y_{gj}^{PM} - Y_{gj}^{MM})/(Y_{gj}^{PM} + Y_{gj}^{MM})$ is performed on the $J_g$ observations to obtain a p-value[1]. The default behavior of MAS 5.0 is to assign a P, M or A call to a p-value smaller than 0.4, between 0.4 and 0.6, and bigger than 0.6 respectively. Liu et al. (2002) demonstrate that the algorithm works relatively well in practice. However, in this section we demonstrate that our framework can be used to generate similar detection calls with almost half of the number of probes.

4.1.2. *Our solution.* Empirical results do not support the assumptions $\mu_{gij}^{PM} = \mu_{gij}^{MM}$ and $S_{gij}^{MM} = 0$. There is strong evidence that $S_{gij}^{MM} > 0$ for many probes [Irizarry et al. (2003a)] and that $\mu_{gij}^{PM} \neq \mu_{gij}^{MM}$ [Naef and Magnasco (2003) and Wu et al. (2004)]. Notice that if we can not use the MM probes, then we need to have probe-specific information about $\mu_{gij}^{PM}$. Wu et al. (2004) describe methodology for estimating $\mu_{gij}^{h}$ using probe sequence information. These authors present two strategies: one uses the MM intensities and the other one does not. In both instances, $\mu_{gij}^{h}$ is estimated by fitting a 15-parameter model to hundreds of thousands of probes, thus, it is estimated with enough precision to consider it known in this application. If we treat $\mu_{gij}^{h}$ as constant, then testing the null hypothesis without using the MM probes is straight forward. Notice that GeneChip arrays include one MM for each PM, thus, PM-only arrays can represent twice as many genes at the same price or represent the same genes at half-price. We therefore refer to our approach without using the MMs as the *half-price* procedure. Notice that the commercial arrays created by NimbleGen [Singh-Gasson et al. (1999)] and the new exon arrays from Affymetrix do not include MM probes. The half-price procedure will permit users of these arrays to perform detection calls.

---

[1] MAS 5.0 software tests the null hypothesis that median($R_g$) = $\tau$ versus the alternative hypothesis that median($R_g$) > $\tau$ for a positive constant $\tau$. The default is $\tau = 0.015$.

4.1.3. *Assessment data.* To compare the two approaches, we used Affymetrix's spike-in experiment on the HG-U133 platform. This experiment is similar to the one described in Irizarry et al. (2003b) and Cope et al. (2004). In this experiment transcripts from 42 genes were artificially added or *spiked-in* to a complex cRNA target at 14 different concentrations ranging from 0 to 512 picoMolar. Fourteen different mixtures were formed by varying the concentrations following a Latin-square design. Three replicates of these mixtures were formed and hybridized to 42 GeneChip arrays of the same type. The 42 spiked-in genes were known not to be present in the original cRNA target, thus, if their spike-in concentration was 0, then the correct detection call is A. For all other concentrations the correct call is obviously P.

4.1.4. *Results.* Figure 2 compares the results obtained using Affymetrix's default procedure and our proposed approach. Figure 2A shows ROC curves for detecting target presence for genes, at various concentrations ranging from 0.125 to 2 picoMolar, using our approach (solid lines) using MM data or MAS 5.0 (dashed lines). Using both the PM and MM measurements as MAS 5.0, our approach tops the result from MAS 5.0. Figure 2B compares the result from the half-price approach(solid line) with MAS 5.0 (dotted lines). Using nearly half the information, the half-price approach achieves similar sensitivity and specificity as MAS 5.0, suggesting this is a useful alternative when no MM measurements are available. In fact, the Affymetrix Exon arrays have adopted a design without MM probes. Another advantage of our approach is that it can easily be extended to include replicates and compute one $p$-value for each gene under each condition. The Affymetrix algorithm assigns Present/Absent calls to each gene on each array. Common practice to dealing with replicates is to call a gene "present" when the number of present calls exceed an arbitrary proportion of the replicates within a condition.

4.2. *Estimating differential expression.*

4.2.1. *Scientific problem.* In this application we typically have two classes of samples (e.g., experimental and control) and in many cases we have various replicates. We are interested in measuring differential expression for each gene. Currently, the standard approach is to first preprocess the probe-level data and then use statistical procedures developed for gene-level data [Chu, Weir and Wolfinger (2002), Dudoit et al. (2002), Kerr et al. (2002), Kerr, Martin and Churchill (2000), Lee et al. (2000), Lonnstedt and Speed (2002), Newton et al. (2001), Schena et al. (1996), Tusher, Tibshirani and Chu (2001), Wolfinger et al. (2001), Yang et al. (2002)] without consideration of the preprocessing algorithm. In this example we will use data from GeneChip arrays.

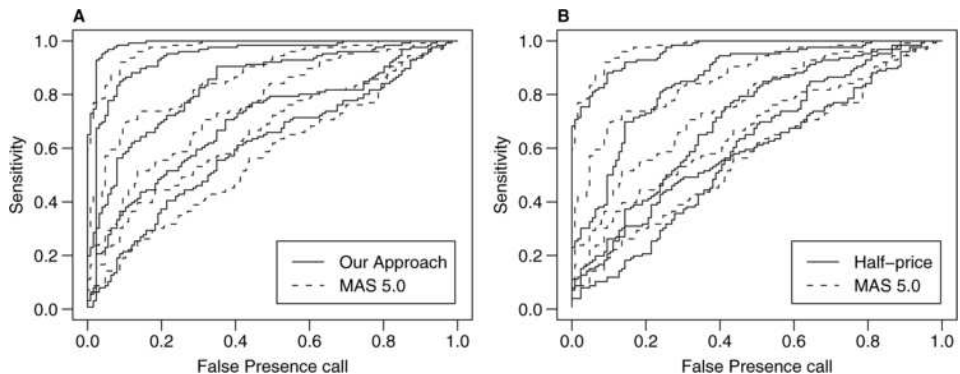

Fig. 2.  (A) *Receiver operating curves for detection calls from MAS* 5.0 (*dashed lines*) *and our approach* (*solid lines*), *when both PM and MM measurements are used. The concentrations of the present genes are* 0.125, 0.25, 0.5, 1 *and* 2 *picoMolar.* (B) *Same as* A *except the half-price approach* (*solid lines*) *is compared to MAS* 5.0 (*dashed lines*).

4.2.2. *Our solution.*  In this context, we quantify differential expression by defining $\theta_{ii} \equiv \beta_{0,g} + \beta_{1,g}X_i$, with $X_i = 1$ if array $i$ was hybridized to the experimental target, and $X_i = 0$ otherwise. The parameter of interest will be $\beta_{1,g}$. For this application we use the MM probes only to estimate $\mu_{gij}$. To do this, we assume that $S_{gij}^{MM} = 0$ which, as seen below, does yield useful results. To reduce the number of parameters needed to represent the probe-specific mean levels, we use probe sequence information as described by Wu et al. (2004). In summary, the $\mu_{gij}^h$ and $\phi_{gij}^h$ are assumed to be linear functions of indicator variables denoting what base (G, C, T or A) is in each position of the probe. We assume the base effect is a smooth function of position and use splines with 5 degrees of freedom to model these functions. These assumptions reduce the number of parameters from hundreds of thousands to less than 20. See Wu et al. (2004) for more details. Other minor assumptions about across and within array correlations are described in the Appendix.

With the specifics of the model in place, we are ready to estimate $\beta_{1,g}$. A possibility is to obtain the MLE along with a standard error for this estimate. Alternatively, we can pose a Bayesian model and obtain posterior distributions of the $\beta_{1,g}$. Wu et al. (2004) demonstrate that normal distribution is a reasonable assumption for $\xi_{gij}$ in model (3). Rocke et al. (2001) show that normal distribution is a good approximation for $\varepsilon_{gij}$ as well. The total intensity of nonspecific and specific binding, $N + S$, is therefore a convolution of two log-normally distributed random variables. This convolution has a complex likelihood and makes it computationally impractical to obtain MLE or Bayesian estimates for each of the thousands of genes on an array. On the other hand, the first two moments of $N_{gij}$ and $S_{gij}$ are easy to compute given the parameters. Therefore, we use generalized estimat-ing equations which rely on simply the first two moments. Details of the implementation are in the Appendix.

Notice that, unlike the modular approach taken by Liu et al. (2006) to propagate the measurement error of expression level summaries into differential expression detection, in this framework, expression level measurements for each array are never calculated. Instead, the parameter of interest is calculated along with a measure of uncertainty that includes the effects of background adjustment, normalization and summarization. We refer to the procedure that leads to an estimate $\hat{\beta}_{1,g}$ and its standard error as the *unified* approach.

4.2.3. *Assessment data.* To demonstrate the utility of the unified procedure, we use data from the spike-in experiment described in Section 4.1.3. Recall that the RNA samples in this experiment were the same in all hybridizations except for the spiked-in genes. The spiked-in genes varied in concentration, within and across arrays. This implies that we can find comparisons of arrays for which only 42 genes are expected to be differentially expressed. Furthermore, for various comparisons, we had three technical triplicates in each group. We choose comparisons of two triplicates for which the expected fold-changes, for most of the spiked-in genes, was 2.

4.2.4. *Results.* Figure 3A shows $\hat{\beta}_{1,g}$ plotted against the average log expression level (taken across the six arrays) for each gene. Notice that this provides similar information to an MA-plot. The blue bars denote pointwise critical values for rejecting the hypothesis that $\beta_{1,g} = 0$ at the 0.01 level. These critical values are computed using the fact that our estimates are asymptotically normal. Nonspiked in genes, which are known not to be differentially expressed, exceeding these bounds are shown with red stars. Spiked-in genes are shown with large purple dots. We add detection call information (described in Section 4.1) for all other points. Yellow represents low $p$-value (present gene), red represents large $p$-value (absent gene). In this case the null hypothesis was that the genes were absent in all six hybridizations. A common approach used by biologists is to filter genes with Affymetrix produced absent calls and then compute fold change estimates. Figure 2 demonstrates that this will result in many false negatives. We propose looking at both fold change estimates and p-values in one plot such as Figure 3A. Because we are adding P/A call information to an MA-plot, we refer to this as an MA-PA plot.

Figure 3A demonstrates that a procedure calling genes differentially expressed when they are outside the critical value bounds performs rather well. Figures 3B compares our results to those obtained with the commonly used approaches, the t-test and the linear models for microarray data (LIMMA) [Smyth (2004)]. LIMMA is one of the most popular procedures for detecting differentially expressed genes among biologists and was designed specifically for this application. This method requires expression-level data, thus, we demonstrate results obtained using two popular preprocessing algorithms: RMA and MAS 5.0 (Affymetrix's default). Figure 3B shows average ROC curves for the five procedures obtained from 14 *three versus three* comparisons. To imitate real data, we excluded comparisons with unrealistically large nominal fold changes and with high nominal concentrations. Specifically, only comparisons with nominal fold changes of 2 and nominal concentrations smaller than 4 picoMolar were included. The ROC curve demonstrates that our procedure performs better than the modular approaches. Figure 4 shows log fold change estimates obtained with the three procedures and demonstrates that our unified approach provides estimates with less bias.

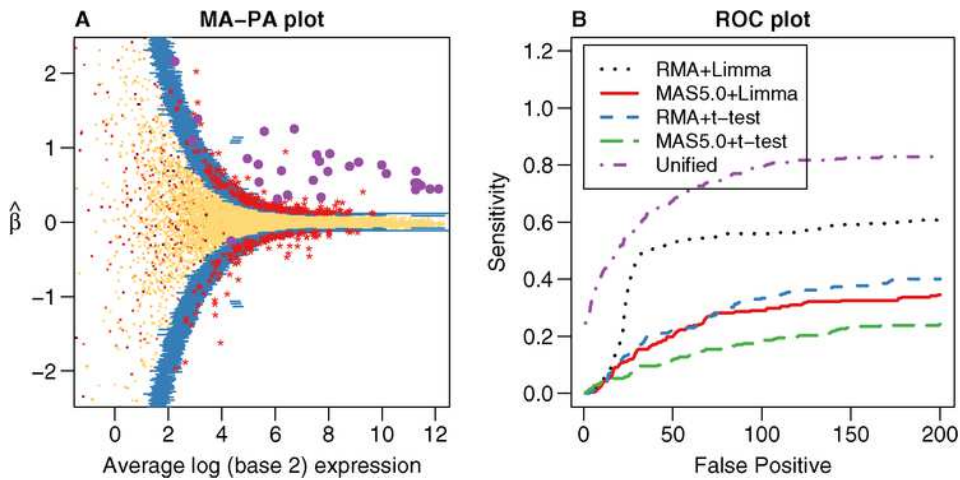

FIG. 3.  (A) $\hat{\beta}_{1,g}$ plotted against $(\hat{\beta}_{1,g} + \hat{\beta}_{0,g})/2$. The color of each dot represents the p-value. Yellow represents low p-value (present genes), red presents large p-value (absent genes). The blue bars mark $\phi(0.995)\hat{SE}$ and $\phi(0.005)\hat{SE}$, where $\phi(\cdot)$ is the cumulative density function of Normal(0,1). Spiked-in genes are labeled as big purple points. (B) Averaged receiver operating curves from 14 comparisons of 2-condition with 3-replicates each from the GeneChip spike-in experiment.

Figure 4 demonstrates that for low nominal concentrations the unified approach estimates have larger variances. However, model based standard error estimates account for this fact. Figure 5 plots our model-based standard error estimate against observed average log intensity for each gene. We also plot sample standard deviations, calculated from 42 arrays, of $\hat{\beta}_{1,g}$ for various strata of the average log intensity. The model-based standard errors, based on three replicates, are very close to the sample standard errors. Notice the strong dependence of both standard error estimates on the average log intensity. This dependence is predicted by our model. Equation (4) in the

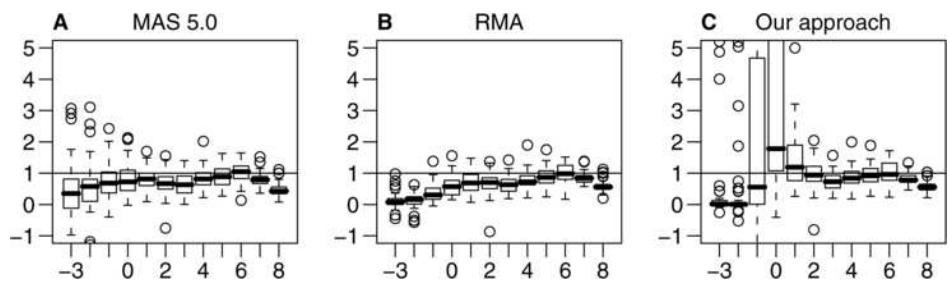

Fig. 4. (A) *Estimated differential expression for genes with nominal fold change of* 2 *obtained with MAS* 5.0. *The x-axis shows the lower of the two nominal concentrations involved in the comparison.* (B) *As* (A) *but using RMA.* (C) *As* (A) *but with our estimate* $\hat{\beta}_{1,g}$.

Appendix shows that the standard error is proportional to the inverse of $\mathrm{E}[S]$. This provides a plausible alternate explanation to the common claim made by many biologists that the high variation observed for low abundance genes is a biological reality. Our calculations show that the high variation is due to the statistical manipulations needed to correct for background.

4.3. *Identification of synthetic lethality and fitness defects in yeast mutants.*

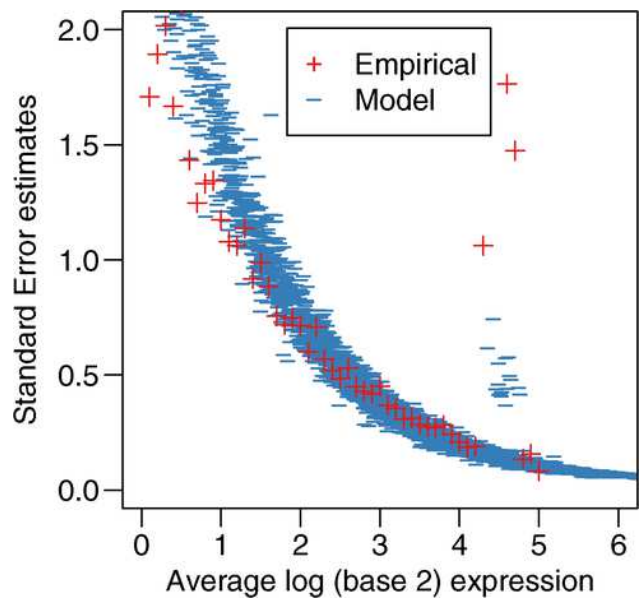

Fig. 5. *Model based (blue bars) and empirical standard deviations (red crosses) of* $\hat{\beta}_{1,g}$ *as described in the text.*

4.3.1. *Scientific problem.* The Yeast Deletion strain collection was created by an international consortium of yeast geneticists [Giaeve et al. (2002)] and is an invaluable resource for genetics research. For each of the 6000+ genes in the yeast genome, a *mutant* yeast strain was created missing that gene. Some genes are essential and, thus, the mutants are not viable. Two unique DNA tags were incorporated into the genome of each mutant strain. Recently, two-channel microarray technology has been developed containing the necessary probes to detect the tags [Yang et al. (2005)]. Thus, Microarray hybridization can be used to measure the *representation* of each mutant in a complex mixture of many different mutants.

A new collection of mutant yeast that are missing two genes is being created. Of interest is to find pairs of nonessential genes for which removing both causes lethality or *fitness to grow* defects. In a typical hybridization, various tags will be missing in the experimental target, these represent dead yeast, and present in the control target, these represent live yeast. Mutants with fitness defects will be under-represented in the experimental target. The task is to identify these tags using the microarray data.

4.3.2. *Our solution.* Because of financial constrains (for each of the 4000+ nonessential genes we need a hybridization), we will typically have only one array $I = 1$ per *query gene.* As mentioned, two tags are used to represent each gene; thus, we have two probes per mutant, that is, $J = 2$ for all $g$. Because the yeast mutants are either dead or alive, we will model $\theta_g^h$ with a two component mixture distribution. One component will represent the dead mutant, that is, $S_{gj}^h = 0$ for $h = R, G$, the other will represent the live mutants. Figure 6A plots a density estimate of log intensities for both R and G channels and clearly shows both alive and dead components. This figure motivates the assumption that $\theta_g^h$ follows a normal distribution for the alive mutants. Furthermore, the figures also supports our claim that the normal assumption for $\xi_{gij}^h$ is useful.

Once the model is fitted under these assumptions, we are ready to provide useful summaries. To quantify the evidence for a gene being dead in the experimental target and alive in the control, we compute a likelihood ratio comparing a model where $Y^R$ and $Y^G$ come from different mixture components to a model where they come from the same. For mutants that appear to be alive in both cultures we can estimate the difference in representation $\log(\theta_g^G) - \log(\theta_g^R)$.

4.3.3. *Assessment data.* One mixture of yeast DNA was split into two halves, and into each half DNA from a few selected mutants were spiked in with known concentration ratio. The concentrations were chosen so that (1) some mutants were not represented in the experimental pool and represented in the control and (2) some mutants had known fold changes in representation when comparing both samples. The spike-in material was introduced into the hybridization mixture in three different concentration groups (high, medium and low). See Peyser et al. (2005) for more details.

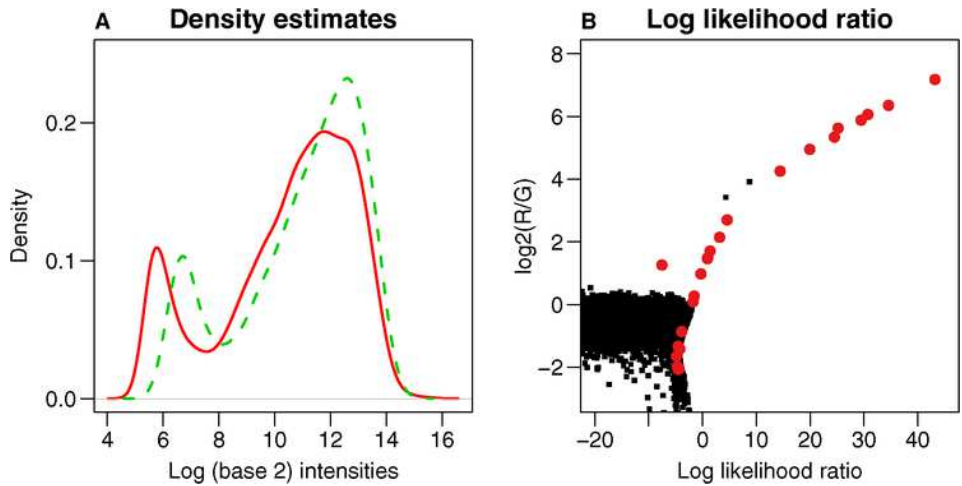

Fig. 6. (A) *Density estimates of log intensities from the Green (dashed green line) and Red channels (solid red line). (B) Log intensity ratios plotted against the log likelihood ratios described in the text.*

4.3.4. *Results.* In Figure 6B we show the log likelihood ratios of mutants that had the same representation (imitating alive/alive or dead/dead) or were spiked-in only in one sample (imitating dead/alive) plotted against the naive log-ratio statistic. This figure shows that the log likelihood ratio statistic clearly discriminates the dead/alive mutants from the rest. Various number of these genes would not have been detected had we used the log ratio.

In Figure 7 we show box-plots of the MLE of $\log(\theta_g^G) - \log(\theta_g^R)$ for the genes that were spiked in to be differentially represented stratified by concentration groups. In this figure we also show estimates obtained using two standard preprocessing procedures. The first is what we refer to as the default procedure which background corrects using the direct estimates of background noise and normalizes by the log ratio medians. The second is the approach proposed by [Dudoit et al. (2002)]. The figure demonstrates our estimation procedure offers an improvement in accuracy and precision over the other two. As in the previous example, the uncertainty introduced by the background adjustment and normalization can be included with our result.

**5. Discussion.** We have presented a general statistical framework for the analysis of microarray data. The advantages of our framework, as any model-based inference, depends on the validity of our assumptions. However, we believe it to be a general enough framework for it to be relevant in many microarray applications, and targeted enough to be useful in practice. We have demonstrated the flexibility of our proposal with three examples from three very different applications and two different platforms. These examples are not intended to be final solutions to the specific problems we presented but rather examples of the adaptability of the proposed framework. An immense amount of work has been published in the statistics literature for both preprocessing and higher-level analyses of microarray data. Our hope is that our work will serve as a basic infrastructure that will permit the integration of these two bodies of work.

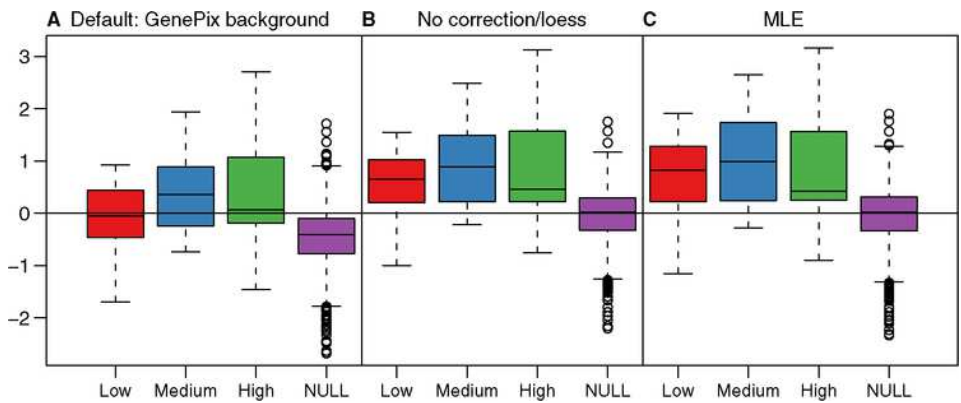

Fig. 7. (A) *Box-plot of log fold change estimates using the default preprocessing algorithm for the low, medium and high concentration groups. The fourth box-plot shows the log fold change estimates of genes that were not spiked-in.* (B) *As* (A) *but with a popular alternative preprocessing algorithm.* (C) *As* (A) *but with our model-based estimate.*

The first step in building a unified model is to describe deterministic effects and all sources of random variation. This allows a complete description of the data structure. However, keeping all these error terms in the model often increases the computational cost. In many practical situations, empirical evidence suggests feasible simplifications. For example, we have observed that the variance due to optical noise is mostly negligible compared to the variance due to nonspecific binding background. Therefore, in the examples we have presented in this paper, we have assumed that $O$ is a constant. In general, we have assumed that the random effects $\xi$ and $\varepsilon$ are correlated among repeated measures. The strength of these correlations depends on the relative magnitude of between and within probe variance. In cases where biological variation is large, the within probe variances across arrays are higher and these correlations will be smaller than what we encounter in technical replicates, as in the Latin-square dataset. In our example for estimating differential expression, we do not differentiate the sources of cross array variance and let the data determine plug-in estimates of the correlation coefficients. Rattray et al. (2006) take a different approach by propagating the uncertainty of gene expression measures obtained in preprocessing procedures into a higher-level model, adding biological variance for each gene. When biological variance is dominant, this approach will probably generate results similar to that from a unified model. However, when the uncertainty from preprocessing is large, the resulting error terms can be highly correlated because the same probes are used on each array. Existing error propagating approaches ignore this fact and will result in overestimates of the gene specific variance across samples. Last, we want to point out that both approaches to integrating the errors from preprocessing and from biological replicates are expected to make a difference mostly in situations with limited replicates. When there is sufficient replication, the benefit of integrated approaches is limited and may not be worth the complexity of computation.

We view our proposal as a first step toward a general framework for microarray probe-level data analysis. The results obtained in three independent applications perform well when compared to procedures that have been fine tuned for that specific applications. Other researchers have also proposed models based on this framework on other applications of microarrays. Wang et al. (2006a) propose a model for estimating genome-wide copy number that is essentially built upon this framework. Meyer et al. (2006) adapt the model from Wu et al. (2004) for ChIP-chip experiment to identify transcription factor binding sites. Results in these applications are encouraging, but further improvements are possible. Below we describe specific aspects that we plan to improve and develop in the future.

Notice that we have purposely left image processing out of our framework. Our experience has been that current default image processing software is reliable and that alternative algorithms offer small or no advantages. In our opinion the added complexity required to model pixel level data is not merited. However, as the feature sizes on the array become smaller, and less pixels are available per feature, it is possible that expanding the framework to describe pixel level data is worth the effort.

Normalization is one of the most controversial topics among users of the technology. Our approach of representing the need for normalization with one parameter is admittedly a simplification. Our current model is likely to improve by considering more elaborate approaches. Furthermore, one of the current limitations of microarray technology is that normalization techniques are based on assumptions related to the number of genes that are changing or the distributions of gene expression across samples. For the majority of experiments these are useful assumptions, making these methods an invaluable resource for improving the quality of microarray data. However, the number of experiments in which these assumptions are not adequate is growing. New normalization schemes will need to be developed. Although offering a specific solution to this problem is outside the scope of this paper, our framework is general enough to adjust to these type of data.

Notice that in model (3), if we remove background, intensity grows linearly with amount of target. Various groups, for example, Hekstra et al. (2003) have noticed that probe hybridization can be modeled using the Langmuir or Hill equations. These models predict a nonlinear behavior. In particular, they predict the observed phenomenon that changes are attenuated for high-abundance genes. Furthermore, various groups, for example, Cludin et al. (2001), have noticed that scanner saturation has a similar effect. In our experience this effect affects a very small proportion of genes [Irizarry, Wu and Jaffee (2006)]. However, adapting our model to account for these effects is possible.

Finally, high-level analyses certainly include more than the examples we give in this paper. For example, gene networks describing the interaction and regulation among genes are often more meaningful than inference on single genes. Studying gene networks using microarrays often relies on the ability to estimate the co-regulation of gene expression. The Pearson correlation of gene expression is often used as a measure of similarity [Getz, Levine and Domany (2000)]. However, in cases with only a few biological samples, gene-level expression summaries do not yield reliable estimates of correlation coefficients. Using the probe level model, we can take advantage of the fact that each probe within a probeset respond to the same biological variation. Estimating correlation of gene expression from the two sets of probe level data can increase the efficiency over the gene-level expression measures.

## APPENDIX

**A.1. Generalized estimating equations for GeneChip spike-in experiment.**
To define the model, we let $\mathbf{Y}_{gj} = (Y_{g1j}, Y_{g2j}, \ldots, Y_{gij}, \ldots, Y_{gIj})'$ denote the vector of PM intensities for probe $j$ across the samples $i = 1, \ldots, I$. Similarly, $\mathbf{N}_{gj}$, $\mathbf{S}_{gj}$, $\boldsymbol{\xi}_{gi}$, $\boldsymbol{\varepsilon}_{gj}$ denote the vectors for probe $j$ across samples corresponding to the definition in model (3), $\boldsymbol{\nu} = (\nu_1, \nu_2, \ldots, \nu_i, \ldots, \nu_I)'$ and $\boldsymbol{\xi}_{gj} = (\phi_{gj}, \phi_{gj}, \ldots)'$. We ignore the variance in optical noise and, as explained in the next section, adjust for it by subtracting the minimal intensity on each array. We write the optical-noise-adjusted intensities as

$$\mathbf{Y}_{gj} = \mathbf{N}_{gj} + \mathbf{S}_{gj}$$
$$= \exp\{\boldsymbol{\mu}_{gj} + \boldsymbol{\xi}_{gj}\} + \exp\{\boldsymbol{\nu} + \boldsymbol{\xi}_{gj} + \mathbf{X}^T\boldsymbol{\theta} + \boldsymbol{\varepsilon}_{gj}\}.$$

Here $\boldsymbol{\theta} = (\theta_1, \theta_2)'$ is the vector of the log scale expression in the two conditions, $\mathbf{X}$ is the design matrix.

We compute plug-in estimators for $\boldsymbol{\mu}$, $\phi_{gj}$ and $\boldsymbol{\nu}$ using data from the entire array as described in the next section. We add constraints that the mean of $\nu$s and $\phi$ of average affinity are 0 such that $\boldsymbol{\theta}$ is identified. We use probe sequence information to predict $\mu_{gj}$. However, the probe effects are not completely accounted for by that linear function base effect. Therefore, considering the same probes are used across arrays, we allow the random effects to be correlated: $\text{var}(\boldsymbol{\xi}_{gj}) = \Sigma^N$, where $\Sigma^N_{ii} = \sigma^2_N$ and $\Sigma^N_{ii'} = \rho_N \sigma^2_N$ for $i \neq i'$. $\text{var}(\boldsymbol{\varepsilon}_{gj}) = \Sigma^S$, where $\Sigma^S_{ii} = \sigma^2_S$ and $\Sigma^S_{ii'} = \rho_S \sigma^2_S$ for $i \neq i'$. The $N$ and $S$ notation denote the nonspecific and specific components. We assume $\boldsymbol{\xi}$ and $\boldsymbol{\varepsilon}$ follow normal distribution and the mean of variance of $Y_{gj}$ is determined accordingly.

We then estimate $\boldsymbol{\theta}$ for each gene $g$ using the following generalized estimating equation:

$$\frac{1}{J}\sum_{j=1}^{J}\mathbf{A}_{gj}(\boldsymbol{\theta})(\mathbf{y}_{gj} - \mathrm{E}_{\theta}[\mathbf{Y}_{gj}]) = 0,$$

where $\mathbf{A}_{gj}(\theta) = (\frac{\partial \mathrm{E}_{\theta}[Y_{gj}]}{\partial \boldsymbol{\theta}})' \mathbf{V}_0^{-1}$ and $\mathbf{V}_0$ is a diagonal working covariance matrix.

The asymptotic variance of $\hat{\boldsymbol{\theta}}$ is

$$\mathbf{D}^{-1}\boldsymbol{\Omega}\mathbf{D}^{-1\prime},$$

where

$$\mathbf{D} = \mathrm{E}\left\{\mathbf{A}_{gj}(\boldsymbol{\theta}_0)\frac{\partial \mathrm{E}_{\theta_0}[Y_{gj}]}{\partial \boldsymbol{\theta}}\right\},$$

$$\boldsymbol{\Omega} = \mathrm{E}\{\mathbf{A}_{gj}(\boldsymbol{\theta}_0)\,\text{var}_{\boldsymbol{\theta}_0}(Y_{gj})\mathbf{A}_{gj}(\boldsymbol{\theta}_0)'\}.$$

We estimate $\mathbf{D}$ and $\boldsymbol{\Omega}$ with

$$\hat{\mathbf{D}} = \frac{1}{J}\sum \mathbf{A}_{gj}(\hat{\boldsymbol{\theta}})\frac{\partial \mathrm{E}_{\hat{\theta}}[Y_{gj}]}{\partial \boldsymbol{\theta}},$$

$$\hat{\boldsymbol{\Omega}} = \frac{1}{J}\mathbf{A}_{gj}(\hat{\boldsymbol{\theta}})\,\text{var}_{\hat{\theta}}(Y_{gj})\mathbf{A}_{gj}(\hat{\boldsymbol{\theta}})'.$$

Notice the averaging is taken over the probes and the sample size here is the number of probes. Although the number of probes in a probe set is not very large (11–20 and typically 16 in GeneChip arrays), Figure 5 shows that the estimated variance based on this asymptotic result fits the observed variance quite well. We believe this is explained by the fact that, although marginal distribution of intensities is right skewed, condition on probe, the distribution of intensities across arrays is approximately normal. This results in a fast convergence of the sandwich estimator of the variance.

An interesting application of these results is that we can illustrate the relationship between the asymptotic variance and the magnitude of $\theta$. We consider the simplest null case where $\boldsymbol{\theta} = (\theta, \theta, \ldots, \theta)'$, $\nu_i = 0$ and all probes in this probeset have the same probe effect. We consider a simple $k$-control/$k$-treatment comparison, therefore, $\mathbf{X}^T$ is the matrix

$$\begin{bmatrix} 1 & 1 & \cdots & 0 & 0 & \cdots \\ 0 & 0 & \cdots & 1 & 1 & \cdots \end{bmatrix}.$$

To simplify notation, we use $\gamma_1 \equiv E[N_{gij}]$, $\gamma_2 \equiv E[S_{gij}]$, $V \equiv \mathrm{var}(Y_{gij})$, $W \equiv \mathbf{cov}(Y_{gij}, Y_{gi'j})$. The normal assumption about $\epsilon$ implies $\gamma_2 = e^{\phi + \theta + \sigma_S^2/2}$ and $\frac{\partial \gamma_2}{\partial \theta} = \gamma_2$. Therefore,

$$A = \frac{\gamma_2}{V} \mathbf{X}^T,$$

$$D = \frac{k\gamma_2^2}{V} \mathbf{I}_{2 \times 2},$$

$$\Omega = \frac{\gamma_2^2}{V^2} \begin{bmatrix} k[V + (k-1)W] & k^2 W \\ k^2 W & k[V + (k-1)W] \end{bmatrix}.$$

The asymptotic variance of $\hat{\theta}$ is then

$$D^{-1}\Omega D^{-1\prime} \propto \gamma_2^{-2} \begin{bmatrix} k[V + (k-1)W] & k^2 W \\ k^2 W & k[V + (k-1)W] \end{bmatrix}$$

and the variance of $\hat{\theta}_1 - \hat{\theta}_2 \propto (V - W)/\gamma_2^2$. Using the normal assumption again, we have $V = \gamma_1^2(e^{\sigma_N^2} - 1) + \gamma_2^2(e^{\sigma_S^2} - 1)$ and $W = \gamma_1^2(e^{\rho_N \sigma_N^2} - 1) + \gamma_2^2(e^{\rho_S \sigma_S^2} - 1)$. This implies

$$(4) \qquad \mathrm{var}(\hat{\theta}_1 - \hat{\theta}_2) \propto \frac{\gamma_1^2(e^{\sigma_N^2} - e^{\rho_N \sigma_N^2}) + \gamma_2^2(e^{\sigma_S^2} - e^{\rho_S \sigma_S^2})}{\gamma_2^2},$$

which predicts that the variance of estimated differential expression converges to a constant as expression levels increase ($\gamma_2$ increases) and is approximately proportional to $1/S^2$ when $S$ is small.

**A.2. Ad-hoc plug-in estimates.** For our example, we assume $O_{gij}$ is constant and form an estimate $\hat{O}$, using the minimum observed intensity on each array. We do this because the variance of $O_{gij}$ is negligible compared to the variance of $N_{gij}$ [Wu et al. (2004)]. To estimate $\mu_{gij}$, probe *affinities* $\alpha_{gj}$ are computed using probe sequence as described in Wu and Irizarry (2004). We assume that $\mu_{gij}$ is a smooth function $h$ of these affinities, that is, $\mu_{gj} = h(\alpha_{gj})$, and estimate $\mu_{gij}$ through estimating $h$. Specifically, a loess curve is fit to the $\log(Y^{MM} - \hat{O})$ versus $\alpha^{MM}$ scatter plot to obtain $\hat{h}$. The $\mu_{gij}$ are then estimated as $\hat{h}_i(\alpha_{gj}^{PM})$. The residuals from the loess fit are used to estimate the variance of $\xi$, $\sigma_N^2$. To estimate the correlation coefficient $\rho_N$, we identify a subset of probes with $\log(Y^{PM} - \hat{O})$ less than the corresponding $\hat{\mu}$. The target mRNAs of these probes are likely to be absent and $\log(Y^{PM} - \hat{O}) \approx N$. We obtain sample variance of each probe across arrays and use $\hat{\sigma}_{N0}^2$ to denote the mean of these variances. $\rho_N$ is calculated as $(\hat{\sigma}_N^2 - \hat{\sigma}_{N0}^2)/\hat{\sigma}_N^2$.

To estimate $\Sigma^S$, we first identify a subset of probe sets with high expression level such that $\log(PM) \approx \log(S)$. Within each probe set we estimate the sample variance of $\log(PM)$ and use the mean as the estimate of $\sigma_S^2$. Using these probes, we regress $\log(PM)$ on $\alpha^{PM}$ to predict $\phi$ for all probes. To estimate $\rho_S$, we use a similar approach as for $\rho_N$: from a subset of probes with strong signals, we obtain sample variance of each probe across arrays and set $\hat{\sigma}_{S0}^2$ as the mean of those variances. $\rho_S$ is calculated as $(\hat{\sigma}_S^2 - \hat{\sigma}_{S0}^2)/\hat{\sigma}_S^2$. We estimate $\theta$ first under $\nu = 0$. The normalization parameters $\nu$ are then estimated such that the $\hat{\theta}_1 - \hat{\theta}_2$ has weighted mean 0, with the weights from estimated standard errors of $\hat{\theta}_1 - \hat{\theta}_2$.

For genes that are not expressed in at least one condition, $\theta = -\infty$ and the GEE may not converge in obtaining $\hat{\theta}$. For these genes we compute a $p$-value testing the hypothesis that it is absent under both conditions. We compute $p$-value for differential expression for all other genes based on the asymptotic normal distribution.

The code implementing the GEE for GeneChip arrays is made into an R package *uniarray* and is available at http://www.stat.brown.edu/~zwu/software.html.

**Acknowledgments.** The authors thank Magnus Rattray and Xuejun Liu for their help with the *pplr* model.

Department of Community Health
Brown University
121 South Main St. Center for Statistical Sciences
Providence, Rhode Island 02906
USA
E-mail: zhijin_wu@brown.edu

Department of Biostatistics
Johns Hopkins University
615 N. Wolfe St. E3620
Baltimore, Maryland 21205
USA
E-mail: rafa@jhu.edu